\documentclass[fleqn,usenatbib]{mnras}
\usepackage{newtxtext,newtxmath}
\usepackage{booktabs, siunitx, multirow, tabularx, adjustbox}
\usepackage[T1]{fontenc}

\DeclareRobustCommand{\VAN}[3]{#2}
\let\VANthebibliography\thebibliography
\def\thebibliography{\DeclareRobustCommand{\VAN}[3]{##3}\VANthebibliography}

\usepackage{graphicx}	
\usepackage{amsmath}	

\usepackage[normalem]{ulem}

\newcommand\RGMX{\bgroup\markoverwith{\textcolor{cyan}{\rule[0.5ex]{4pt}{1pt}}}\ULon}

\newcommand\ACCX{\bgroup\markoverwith{\textcolor{red}{\rule[0.5ex]{4pt}{1pt}}}\ULon} 

\title[Terrestrial circumbinary planet formation]{
Terrestrial planet formation in a circumbinary disc around a coplanar binary} 

\author[A. C. Childs \& R. G. Martin]{
Anna C. Childs\thanks{E-mail: childsa6@unlv.nevada.edu}
and Rebecca G. Martin
\\
Department of Physics and Astronomy, University of Nevada, Las Vegas, 4505 South Maryland Parkway, Las Vegas, NV 89154, USA\\
}

\date{Accepted XXX. Received YYY; in original form ZZZ}

\pubyear{2021}

\begin{document}
\label{firstpage}
\pagerange{\pageref{firstpage}--\pageref{lastpage}}
\maketitle

\begin{abstract}
With $n$-body simulations, we model  terrestrial circumbinary planet (CBP) formation with an initial surface density profile motivated by  hydrodynamic circumbinary gas disc simulations.  The binary plays an important role in shaping the initial distribution of bodies. After the gas disc has dissipated, the torque from the binary speeds up the planet formation process by promoting body-body interactions but also drives the ejection of planet building material from the system at an early time.  Fewer but more massive planets form around a close binary compared to a single star system. A  sufficiently wide or eccentric binary can prohibit terrestrial planet formation. Eccentric binaries and exterior giant planets exacerbate these effects as they both reduce the radial range of the stable orbits. However, with a large enough stable region, the planets that do form are more massive, more eccentric and more inclined.   The giant planets remain on stable orbits in all our simulations suggesting that giant planets are long-lived in planetary systems once they are formed. 
\end{abstract}




\section{Introduction}

The \textit{Kepler} space telescope has thus far observed 13 circumbinary planets (CBPs) \citep{Doyle2011, Welsh2012, Orosz2012, Schwamb2013, Kostov2013, Kostov2014, Welsh2015, Kostov2016, Orosz2019, Socia2020}, and the \textit{TESS} space telescope has observed one CBP \citep{Kostov2020}. Most of these CBPs are gas giants and all are larger than the terrestrial planets in our solar system. Most of the orbital periods are around five to six times that of their host binary star orbital period and their orbits are coplanar to the binary orbit.  However, the properties of these observed planets are likely a consequence of observational bias and not  representative of the underlying CBP  population \citep{Czekala2019,MartinD2019}. Transiting planets around binaries are difficult to detect because the transit timing variations are larger than the duration of the transit \citep{MartinFabrycky2021}.  Numerous exoplanet surveys have shown that planetary systems around single stars are diverse, but the extent of this diversity for planets around binaries, and their formation pathways remain open questions.

Circumbinary planets are thought to form from a circumbinary gas disc that forms as a result of the star formation process \citep[e.g.][]{Monin2007,Kraus2012,Harris2012}. The dynamics of circumbinary discs are strongly influenced by the torque provided by the central binary \citep[e.g.][]{Artymowicz94,Artymowicz96,Smallwood2020}.  The size of the cavity created by the binary star and the surface density distribution of the disc depends upon the binary separation, eccentricity and inclination \citep[e.g.][]{Artymowicz94,Miranda_2015,Lubow2018}.

The current standard model for planet formation is the core accretion theory \citep{Artymowicz1987, Lissauer1993, Pollack1996}. This describes a process in which particles that are initially within a gas disc begin as dust grains and collide with one another to form larger bodies until stable planetary systems are formed. Like circumstellar planets, CBPs are thought to form through core accretion although the specifics may differ in the early stages of planetesimal formation between circumstellar and circumbinary discs \citep{Bromley2015, Chachan19}.

Unlike circumstellar discs, circumbinary discs experience strong tidal forces that may  inhibit in situ formation of larger planetary building blocks in the inner disc by increasing the relative planetesimal collision velocities \citep[e.g.][]{Moriwaki2004,Scholl2007, Meschiari12,Paardekooper_2012,Silsbee_2015, Marzari2013} and reducing the pebble accretion efficiency \citep[e.g.][]{Pierens20,Penzlin2020}. This has led to the suggestion that CBPs may form in more distant regions where the time-varying potential of the binary is weaker, and then migrate inwards via disc interactions to their observed orbits \citep{Pierens_2008, Bromley2015,Penzlin2020}.  However, the issues with in situ planetesimal formation may be overcome if the protoplanetary disc is sufficiently massive  \citep[e.g.][]{Marzari2000,Martin2013,Meschiari2014,Rafikov2015}. Circumbinary gas discs have a longer disc lifetime and may be more massive than circumstellar discs since the binary torque can reduce the accretion rate on to the stars \citep[e.g.][]{Alexander2012} and this may aid in overcoming the planetesimal formation problems.  Furthermore, \citet{Paardekooper2010} have shown that when fragmentation is accounted for, second generation planetesimals can grow from the fragments of previously collided planetesimals in circumstellar discs of close-in binaries.  We expect an analogous scenario to take place in circumbinary discs.

The late stages of terrestrial planet formation take place after the gas disc has dispersed and are characterised by the gravitational interaction of Moon-sized planetesimals and Mars-sized embryos that form planets \citep{Weidenschilling77, Rafikov_2003, CHAMBERS2001}.  The dynamics that a planetary embryo experiences during this stage  determines the planet's final mass and orbital properties. The late stages of in situ terrestrial planet formation for CBPs in coplanar discs have previously been numerically studied  \citep[e.g.][]{Quintana06,quintana2007terrestrial, Quintana08,Quintana2010,Gong2013,Lines2014,Barbosa2020}. While widely separated, eccentric binaries can inhibit CBP formation, planetary systems can form for a range of binary mass fractions and orbits.

In this paper, we follow the work of \citet{Quintana06} and model the late stages of CBP  formation  for binary systems with different separations and eccentricities using  $n$-body simulations of the late stages of planet formation.  Our study differs in two major ways.  First, we use more realistic initial surface density profiles for the particles that are motivated by hydrodynamical gas disc simulation results.  Secondly, we simulate systems with and without giant planets and consider the effects of external giant perturbers on CBP formation. \citet{Childs19} found that exterior giant planets promote terrestrial planet formation in the inner regions of a circumstellar disc.  \citet{Quintana06} and \cite{quintana2007terrestrial} include giant planets in all their simulations of CBP formation.  We want to understand if the gravitational perturbations from giant planets that promote embryo and planetesimal interactions around single stars can be reproduced by the perturbations from the time-varying potential of the binary. 
 In Section~\ref{sims} we discuss our hydrodynamic simulations and their connection to the setup for our $n$-body simulations.  In Section~\ref{results} we present our results, and in  Section~\ref{concs} we summarise our findings that allow us to make predictions about coplanar planet properties for the so far largely unobserved, terrestrial CBPs.

\section{Initial particle disc setup and methods}
\label{sims}

$N$-body simulations of terrestrial planet formation around a single star typically use an initial surface density profile for the particles that is a power law with radius  \citep[e.g.][]{Hayashi81, Ida04, Miguel08, Miguel08core, Mordasini09}.  This surface density profile is motivated by the observed mass distribution in the solar system  \citep{Weidenschilling77} and since the single star does not exert a torque on the disc, this is a reasonable approximation to a quasi-steady state disc \citep{Pringle81}. In the case of a central binary, the additional torque from the binary means that the initial surface density profile for the particles is not well approximated by a power law close to the binary \citep[e.g.][]{Pringle1991,Gunther2002}.  The profile is also highly dependent on the binary eccentricity \citep{Artymowicz94, Artymowicz96, Lubow_2015,Miranda_2015,Lubow2018,Franchini2019b, Liu19}.
In this Section, we first use smoothed particle hydrodynamic (SPH) simulations to model the surface density profile of a quasi-steady state circumbinary gas disc. We then set up our $n$-body simulations with a surface density profile fit to our SPH results. Finally we discuss the stability limit for test particle orbits and compare it to our initial particle set up.

\subsection{Hydrodynamic circumbinary gas disc simulations}
\label{hydro}

\begin{figure*}
	\includegraphics[width=1.5\columnwidth]{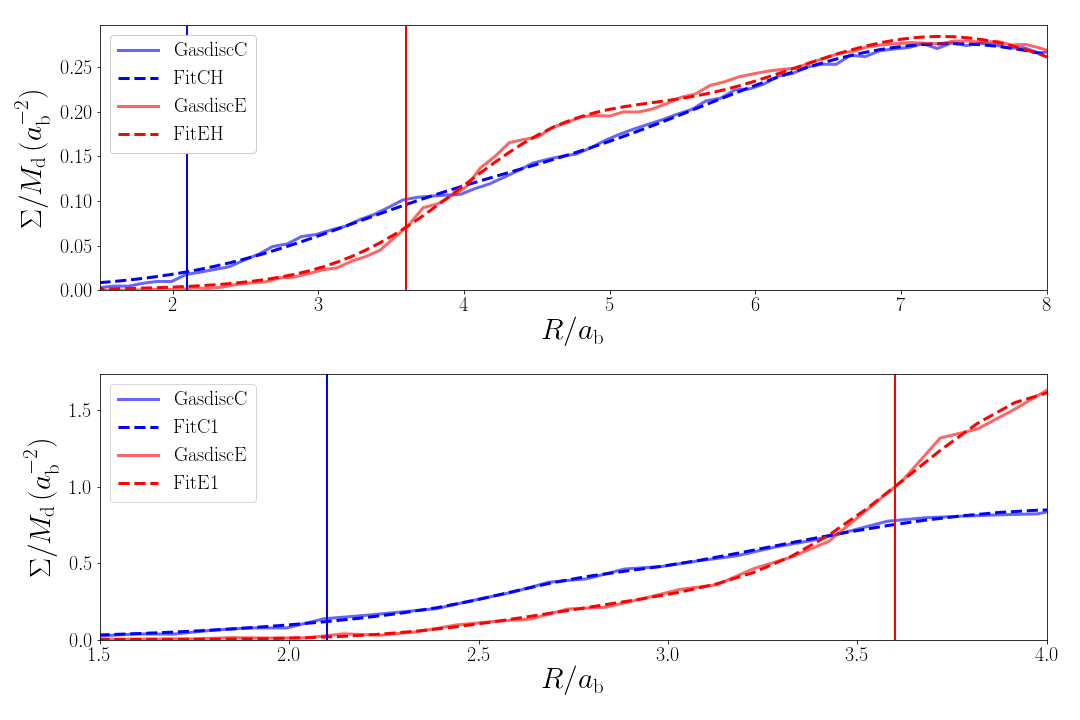}
    \caption{Surface density profiles of the SPH simulations for a circular (GasdiscC) and eccentric binary (GasdiscE), and their analytic fits for the binary models listed in Table \ref{tab:binary_models}.  The data from the SPH results are shown by solid lines and the double Gaussian fits to the SPH data are shown by dashed lines.  We consider the surface density profiles in the ranges [1.5,8.0] $a_{\rm b}$ and [1.5,4.0] $a_{\rm b}$ for binaries separated by $0.5 \, \rm au$ and $1 \, \rm au$ respectively.  The critical stability radius for circular binaries is marked by the blue line and for binaries with $e=0.8$ it is marked in red.} 
    \label{fig:surface_density_profiles}
\end{figure*}

We run two simulations of a circumbinary gas disc around an equal mass  binary ($M_1=M_2=0.5\,M$, where $M$ is the total mass of the binary) with eccentricity $e_{\rm b}$ and semi-major axis $a_{\rm b}$. We use the SPH \citep{Price_2007, Price2012} code {\sc Phantom} \citep{Price10, Lodato10, Price_2018} that has been used extensively for circumbinary discs \citep[e.g.][]{Nixon12,Nixon13,Rocher18,Smallwood_2019, Aly_2020}. The first SPH simulation  is a circular and coplanar orbit binary, GasdiscC. The second simulation is an eccentric ($e_{\rm b}=0.8$) and coplanar orbit, GasdiscE.  

We run each simulation until the inner parts of the disc reach a quasi-steady state,  meaning that the density profile is self-similar in the inner part of the disk at subsequent times. In order to reach a complete steady state we would need to have a steady flow of material added to the outer parts of the disc and to integrate the simulation for a very long time \citep[e.g.][]{Munoz2019}. Since we are interested only in the surface density profile for the disc and the mass scaling is arbitrary, we do not add material to the disc.  The mass of the disc decreases in time as mass falls on to the central binary.

The simulation results are independent of disc mass that we take to be $M_{\rm d}=0.001\,M$. In each case, the disc surface density is initially a power with radius $(\Sigma \propto R^{-3/2})$ between inner radius $R_{\rm in}=6\, a_{\rm b}$ and outer radius $R_{\rm out}=10\,a_{\rm b}$. The initial inner radius is chosen to be far enough away from the binary in all cases that material initially flows inwards so that we can achieve a steady state that does not depend on the initial conditions. The disc spreads inwards and outwards during the simulation.

\begin{table} 
\centering
\setlength{\tabcolsep}{5.0pt}     
\setlength{\cmidrulekern}{0.3em} 
\caption{ Surface density profile fits from the SPH models. All binary orbits are coplanar to the disc. We list which SPH model is used to determine the initial surface density for each binary model, the separation and eccentricity of the binary model, and the outer radius of the fit.}
\begin{tabular}{%
  S[table-format=1.0]
  S[table-format=1.2]
  *{2}{ 
    *{2}{S[table-format=1.3(3)]} 
    S[table-format=1.3]  
  }
}
  \toprule

    {Surface density fit} &  {$a_{\rm b}$(au)} & {$e_{\rm b}$} & {SPH model} & $R_{\rm out}/a_{\rm b}$ \\
    \cmidrule(lr){1-5}                  
    {FitCH} & {$0.5$} & {$0.0$}  & {GasdiscC} & {8}\\
    {FitEH} & {$0.5$} & {$0.8$} & {GasdiscE} & {8}\\
   {FitC1} & {$1.0$} & {$0.0$}& {GasdiscC} & {4}\\ 
   {FitE1} & {$1.0$} & {$0.8$} & {GasdiscE} & {4}\\

  \bottomrule
\end{tabular}
\label{tab:binary_models}
\end{table}

The \citet{Shakura73} viscosity parameter is set to $\alpha=0.01$. The viscosity is implemented by adapting the SPH artificial viscosity according to \citet{Lodato10}.  The disc is locally isothermal with sound speed $c_{\rm s}\propto R^{-3/4}$ and the disc aspect ratio varies weakly with radius as $H/R\propto R^{-1/4}$. This is chosen so that $\alpha$ and the smoothing length $\left<h\right>/H$ are constant with radius \citep{Lodato07}. 
Note that since the SPH simulations are scaled to the binary separation we do not run different simulations for different binary semi-major axis. Instead, we use the same profile (scaled to the binary separation) but truncate it at a different outer radius (relative to the binary separation).  Changing the binary separation  is equivalent to changing the outer radius used in the surface density profiles. The temperature profile of the disc is determined by the disc aspect ratio, $H/R$, that is a fixed in time and scaled to the binary separation such that $H/R=0.05$ at radius $R=6\,a_{\rm b}$.  By scaling the temperature profile in this way, the surface density profile scales with $a_{\rm b}$.

Each simulation contains $500,000$ SPH particles initially. The stars are treated as a sink particles with accretion radii of $0.25\,a_{\rm b}$. The mass and angular momentum of any SPH particle that passes inside the accretion radius is added to the star. We do not include the effects of self-gravity in our calculations. The surface density profiles for the two SPH simulations are shown in the solid lines in Fig.~\ref{fig:surface_density_profiles} at a time of $1000\,P_{\rm orb}$, where $P_{\rm orb}$ is the orbital period of the binary.  The solid lines in the upper and lower panels are the same, the upper panel just extends to larger radius relative to the binary separation. 

In Fig.~\ref{fig:surface_density_profiles} we also show a double Gaussian analytic fit to each profile in the dashed lines. The different fits and their binary setups are shown in Table~\ref{tab:binary_models}. The surface density in all cases becomes very small in $R< 1.5\,a_{\rm b}$ and we fit the distribution down to $R=1.5\,a_{\rm b}$.

We set the outer edge of our disc fits to be $R=4 \,\rm au$ in all cases. This is equivalent to $R=8\,a_{\rm b}$ and $R=4\,a_{\rm b}$ for our two binary separations of $0.5\,\rm au$ and $1\,\rm au$, respectively. Although imposing an outer edge at $4 \, \rm au$ deviates from scaling the disc with binary separation, we do this so that we may include Jupiter and Saturn at fixed orbits in our simulations.  While the SPH simulation surface densities (solid lines) are the same in the upper and lower panels of Fig.~\ref{fig:surface_density_profiles}, just a different radial range,  the fits (the dashed lines) are slightly different.

We expect the late-stage of terrestrial planet formation to take place inside of the snow line radius where water is in a gaseous form \citep[e.g.][]{Lecar2006,Garaud2007,Martin2012}.  Our choice to truncate the outer edge of the disc at $R=4 \,\rm au$ is motivated by previous $n$-body studies. \citet{Quintana2014} used a disc that extended out to $4 \, \rm au$ in order to study the dynamics and radial mixing of volatile rich bodies exterior to the snowline in the solar system.  \citet{Quintana2014} found that late stage water delivery to the inner terrestrial planets most likely originated from volatile rich bodies exterior to the snowline.  Consequently, subsequent work studying terrestrial planet formation has adopted a disc that extends out to $4 \, \rm au$ \citep{Quintana2016, Childs19}.

The initial orbital properties of the orbits of Jupiter and Saturn are the same in all runs that include the giant planets. We set the outer edge of the particle disc to be $4 \, \rm au$ in all runs so we may make a  direct comparison between systems with and without the giant planets. Note that our initial surface density profile has a sharp truncation at the outer edge. This does not account for the shape of the gap in the gas disc that the giant planets would carve out \citep[e.g][]{Lin1986}. Using an approximation formula from \citet{Takeuchi1996} and the viscosity parameter and disc aspect ratio from our SPH simulations we estimate that Jupiter, at its current mass and radius, would open a gap that extends inwards to about  $3.65 \, \rm au$.  While we allow our discs to extend out to $4 \, \rm au$ we expect that the particles that are initially exterior to $3.65 \, \rm au$ become quickly unstable without  significantly affecting the dynamical evolution and architecture of the final planetary systems.  We discuss this further in Section~\ref{unstable}. The outer truncation radius for the planetesimal disc restricts the radial range where terrestrial planets may form and prevents their formation around binaries with wider orbits \citep[e.g.][]{Clanton2013}.

In this work we assume that both  giant planets and terrestrial planets form in situ. However, some solar system formation models allow for migration of the giant planets after their formation. For example, in the Grand Tack model, Jupiter first  migrated inwards  down to an orbital radius of $1.5\,\rm au$ and later  migrated outwards to its current location  \citep{Walsh2011, Raymond_2014}.  This scenario could significantly alter the initial distribution of the particles available for terrestrial planet formation. Since the original locations and migrations of the giant planets are still widely debated topics, we assume in situ formation of the giant planets for simplicity.  In this scenario, the gas profile is a good proxy for the initial distribution of solid bodies after gas dispersal.

\subsection{N-body Simulations}
\label{nbody}

Our $n$-body simulations model the late stages of planet formation after the gas disc has completely dissipated and solid bodies, Moon-size planetesimals up to Mars-size embryos, interact with one another through purely gravitational interactions.  It is through these gravitational interactions of planetesimals and embryos that terrestrial planets form \citep{Kokubo_1996, CHAMBERS2001}.

We use the $n$-body code \texttt{REBOUND} \citep{Rein2012} with the symplectic integrator \texttt{IAS15}  \citep{Rein2015} unless otherwise stated.  \texttt{IAS15} utilises an adaptive time step and we set an initial time step of about $2 \%$ of the binary orbit.  Collisions are resolved by a perfect merging model which always merges particles together if their physical radii are detected to overlap with one another. During the process, the mass and momentum of the particles are conserved.  We define an ejection from the system as a particle that has exceeded a distance of $100\,\rm{au}$.  We remove the particle from the simulation at the time that this criteria is met.

All stars are given a mass of $0.5\,\rm  M_{\odot}$ and a radius of $0.001 \,\rm au$.  We consider different binary models with various values of binary separation ($a_{\rm b}$) and eccentricity ($e_{\rm b}$) for the binary orbit.  The binary parameters for each model are listed in Table \ref{tab:avg_planets}. Each model is a unique set of binary separation ($a_{\rm b}$) and eccentricity ($e_{\rm b}$).  Model names with a C refer to circular orbit binaries and model names with an E refer to eccentric orbit binaries. The remaining part of the model name refers to the binary separation, $a_{\rm b}$, ($1 \, \rm au$ or half an au).  Model names that begin with S are for single star runs which are discussed later on.  The disc particle orbits are measured with respect to the center of mass of the system.  

The range of binary semi-major axes and eccentricities is chosen so that the formation of planet embryos inside of the snow line is possible.  The $n$-body disc is largely motivated by solar system studies.  As a result, we choose a binary whose total mass is $1 \, M_{\odot}$.  Mass ratio distributions of observed binary stars reveal a \textit{twin phenomenon} which refers to an excess of stellar mass ratios near one and so, we choose equal mass stars for our study.  Futhermore, studies with a binary mass ratio close to one focus on spectroscopic binaries with a small separation ($a_{\rm b}<1 \, \rm au$) and so we consider binaries with relatively small separations of $1 \, \rm au$ and $0.5 \, \rm au$ \citep{Lucy1979, Hogeveen:1992wa, Tokovinin2000,refId0, Lucy2006, Pinsonneault_2006, Simon_2009, Kounkel_2019}. The binary orbital plane and gas and particle discs begin close to coplanar which is consistent with previous theoretical studies and most observations of circumbinary debris discs \citep{Kennedy_2012, Foucart_2013, Li_2016}. The eccentricity of the binary is sampled at two extremes of circular ($e=0$) and eccentric ($e=0.8$).

The particle disc we use for our $n$-body studies is adopted from \citet{Quintana2014} and is an extrapolation from the disc used in \citet{CHAMBERS2001} although there is some debate whether embryos may form this close-in to the binary. \citet{Moriwaki2004} found that planetesimals may not form close-in to the binary in a gas-free environment and \citet{Marzari2013} found that even in a gas-rich environment planetesmials have a difficult time growing as binary perturbations grow planetesimal velocities to speeds that are more likely to result in fragmentation rather than accretion.  However, there are mechanisms available that may overcome this barrier to embryo growth interior to the critical stability limit such as second generational growth of planetesimals via fragments \citep{Paardekooper2010}.  Additionally, previous studies of terrestrial planet formation in circumbinary discs consider discs that begin even closer-in to the binary \citep{Quintana06, quintana2007terrestrial}.

To generate the initial particle disc surface profile we use the analytic fits to the results from the SPH simulations described in Section~\ref{hydro} (see Fig.~\ref{fig:surface_density_profiles}).  We then uniformly distribute 26 Mars-sized embryos ($m=0.093\,\rm M_{\oplus}$) and 260 Moon-sized planetesimals ($m=0.0093\,\rm M_{\oplus}$) along the fits between $1.5\,a_{\rm b}$ and $4.0\,\rm au$. The total mass of the planetesimals and embryos is $4.85\,\rm M_{\oplus}$.  We assume that all of the gas has dissipated by this time and our disc now only contains solid bodies.  Assuming a dust-to-gas ratio of 0.01, this dust mass implies an initial gas disc mass of $\sim 0.0015 \, M_{\odot}$ for the inner disc regions.  All bodies begin on nearly circular ($e$ < 0.01) and nearly co-planar orbits ($i<1^{\circ}$).  Body eccentricities and inclinations are uniformly distributed between (0.0,0.01) and ($0^{\circ},1^{\circ}$), respectively.  All other orbital elements are uniformly distributed between $0^{\circ}$ and $360^{\circ}$.  This bi-modal mass distribution and the distribution of orbital elements are extrapolated from the disc used in \citet{CHAMBERS2001}.   \cite{CHAMBERS2001} successfully reproduces the broad characteristics of the solar system and consequently, this disc is used for many $n$-body studies of terrestrial planet formation.  

For the $n-$body simulations we model the inner parts of the disc up to a radius of $R=4\,\rm au$ in all cases. We consider two different binary separations, $a_{\rm b}=0.5\,\rm au$ and $a_{\rm b}=1\,\rm au$.  For the simulations with $a_{\rm b}=0.5\,\rm au$ (CH and EH), we use the fits shown in the upper panel of Fig.~\ref{fig:surface_density_profiles} and for the simulations with $a_{\rm b}=1\,\rm au$ (C1 and E1), we use the fits shown in the lower panel of Fig.~\ref{fig:surface_density_profiles}, as described in Table~\ref{tab:binary_models}.

Unless otherwise stated, we perform 50 runs with giant planets and 50 runs without giant planets for each setup. All runs begin with the same initial conditions for a given model, however we change the random seed generator used for the orbital elements of the planetesimals and embryos in each run.  The systems with giant planets include Jupiter and Saturn at their current orbit and mass.  The Jupiter planet has the initial properties of mass $m = 317.7 \, M_{\oplus}$, semi-major axis $a =5.20349 \, \rm au$, eccentricity $e=0.048381$, and inclination $i=0.365^{\circ}$, and the Saturn planet has $m = 95.1 \, M_{\oplus}$, $a =9.54309 \, \rm au$, $e=0.052519$, and $i=0.8892^{\circ}$. The runs that include Jupiter and Saturn are denoted with the subscript JS. Runs without a subscript do not include Jupiter and Saturn.

To help us identify what effects are caused by the binary we also perform simulations around a single $1\, \rm M_{\odot}$ star using the disc from the CH model and also the disc from the C1 model.  We refer to the runs using the CH disc around a single star as SH and to the runs with the C1 disc as S1.  We integrate 50 runs for both SH and S1 models. We note that the single star simulations presented here are not supposed to be a model of planet formation around a single star since we use the surface density profile of a circumbinary disc. They are simply to enable us to disentangle the binary effect on the planet formation process.

All bodies, excluding the stars, are given an initial density of $3\,\rm  g\,cm^{-3}$. Because \texttt{IAS15} is a high accuracy integrator, in order to reduce computation time we apply an expansion factor to the particle radii of the planetesimals and embryos.  We expand their radius by a factor $f=25$ times their initial radius.  The use of an expansion factor in $n$-body studies was shown by \cite{Kokubo_1996, Kokubo_2002} to not have a significant effect on the evolution of planets other than reducing the timescale of planet formation provided that the velocity dispersion of the bodies is not dominated by gravitational scattering.
Although previous studies mostly use an expansion factor up to about $f=6$, these studies use collision models that allow for inelastic bouncing and/or fragmentation which will significantly affect the gravitational scattering of bodies \citep[e.g.][]{Leinhardt2005,Bonsor2015}.  Since we use a simple merging model, where particles always merge when their physical radii come in contact, we are able to use a larger expansion factor.  Using only perfect merging, \citet{Kokubo_2002} experiment with $f=10$ in $n$-body simulations modeling planetesimal growth and find similar results as their simulations with $f=6$.  After short term experiments with $f=5,10,20,25,100$, we chose the smallest expansion factor that yielded a reasonable simulation runtime. In Section~\ref{ef} we show some convergence tests with different expansion factors.

Aside from the convergence tests, we apply the same expansion factor to all systems and anticipate that the contributions of the expansion factor will have the same effect on all systems to a reasonable degree.  We expect the differences that arise between systems is mainly a result of differences in surface density profiles, binary orbital parameters and the presence or absence of giant planets.

We integrate all our systems with $f=25$ for $7 \, \rm Myr$.  Terrestrial planet formation happens on timescales much longer, up to hundreds of millions of years however, we artificially inflate the particle radii which allows us to identify trends in planet formation pathways after a much shorter integration time as it corresponds to an \textit{effective} timescale in excess of typical terrestrial planet formation timescales.

\subsection{Critical stability limit for a particle}
\label{ac}

Strong perturbations from a central binary clear out planet orbits in the inner regions of the disc drastically lowering the probability of particles existing there \citep[e.g.][]{Holman99,Chen2020}.  An analytical theory for stable circumbinary orbits has been put forth by \citet{Lee06} and \citet{Leung13}, based on the restricted three-body problem \citep{Szebehely67, Murray2000}.  This theory has been tested via $n$-body simulations by \citet{Bromley2015} and \citet{Mason15}.  
The radial stability limit, $a_{\rm c}$, for a coplanar planet is the innermost stable orbit that a planet can reside on around a binary with a given eccentricity, $e_{\rm b}$, mass fraction, $\mu$, and orbital separation, $a_{\rm b}$.  Empirical fits by \citet{Holman99}, improved on by \citet{Quarles_2018}, find that the critical radius, or stability limit, for a binary with a given mass ratio, separation and eccentricity is given by
\begin{equation}\label{eq:a_c}
\begin{aligned}
    a_{\rm c}/a_{\rm b} = 1.48+3.92 e_{\rm b} -1.41 e_{\rm b}^2 + 5.14\mu + \\ 0.33 e_{\rm b}\mu - 7.95 \mu^2 - 4.89 e_{\rm b}^2 \mu^2,
    \end{aligned}
\end{equation}
where
\begin{equation}
    \mu = \frac{M_{\rm s}}{M_{\rm s}+M_{\rm p}},
\end{equation}
 $M_{\rm s}$ is the mass of the secondary star and $M_{\rm p}$ is the mass of the primary star \citep[see also][]{Bromley2015,Quarles_2018,Chen2020}. It should be noted that Equation~(\ref{eq:a_c})  assumes that the planet is coplanar to the binary orbit. 

For an equal mass binary, the stability limit for a circular binary is $a_{\rm c}/a_{\rm b}=2.1$ and for a binary with an eccentricity of 0.8, it is $a_{\rm c}/a_{\rm b}=3.6$. The vertical lines in Fig.~\ref{fig:surface_density_profiles} show the location of critical stability radius compared to our disc surface density profiles. The gas disc is stable closer to the binary than the test particle stability limit. The rings in the gas disc communicate with each other through pressure leading to a stabilising effect. Since we truncate the outer disc edge at $4\,\rm au$ in all cases,  the wider binary separation simulations have a larger fraction of mass initially in $R<a_{\rm c}$. However, once the gas disc has dissipated, particles inside of the critical stability limit become unstable as the critical stability limit is a prediction of stability for solid bodies in the absence of gas.

Similarly to \citet{Quintana06}, our particle discs can begin with  particles that are interior to the critical particle stability limit. We note that it may be difficult for planetesimals to form close to the binary.    While there does remain uncertainty in the early stages of planetesimal formation in the inner regions, we adopt a particle disc with the same surface density profile as the gas disc.  If stable gas is able to grow and harbor embryos and dissipate on a short timescale, then the gas profile is a good proxy for the initial location of the embryos. In Section~\ref{unstable} we consider the effect of these initially unstable particles. 

\section{Results}
\label{results}

In this Section we examine the results of our $n$-body simulations. We first show that the orbital evolution of the giant planets, Jupiter and Saturn, are not affected by presence of the central binary stars. Next, we consider the progress of our simulations in terms of the amount of material that is available for planet formation in time and then we look at the properties of the resulting circumbinary planetary systems.  Finally, we discuss the expansion factor convergence tests and the effect of particles that begin inside of the critical particle stability limit.

\subsection{Giant planet orbital evolution}

Most of the observed CBPs are gas giants. Although this is most likely the result of observational bias, \citet{Armstrong2014} used debiasing processes on observational data to predict the occurrence rate of giant planets and found that giant circumbinary planets appear to be as common as those orbiting single stars.  Because of their high occurrence rate and strong influence on planet formation, we include Jupiter and Saturn in some of our simulations. Assuming that the giant planets have the  orbital properties of Jupiter and Saturn is a reasonable assumption since we expect giant planets to form outside of the snow line radius \citep[e.g.][]{Hayashi81,Kennedy2008,Martin2013b}.

\begin{figure*}
	\includegraphics[width=2\columnwidth]{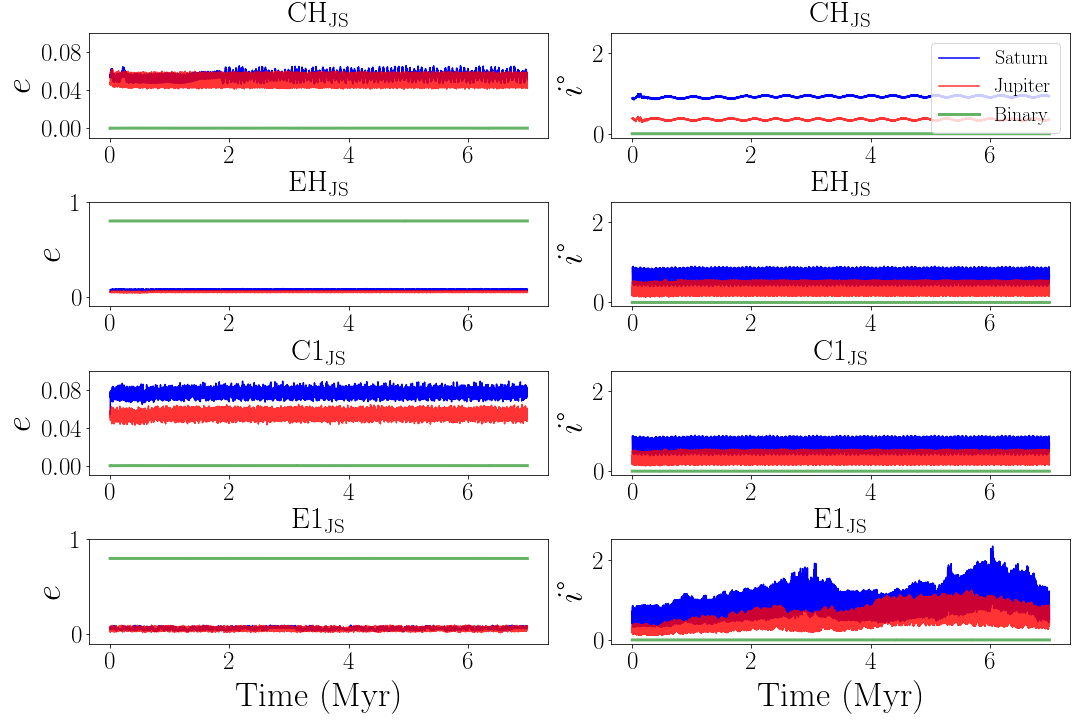}
    \caption{Eccentricity and inclination evolution of the binary, Jupiter and Saturn orbits.  The data shown is from one randomly chosen run for each model.  We find similar behavior of the larger bodies in all runs for a given model. Giant planets remain on stable orbits in all of the simulations although the amplitudes of their oscillations increase slightly with binary separation and eccentricity.}
    \label{fig:big_body_eccs}
\end{figure*}

In all of our giant planet runs, Jupiter and Saturn remain on stable orbits and are never ejected even though they orbit a binary star.  Figure \ref{fig:big_body_eccs} shows the eccentricity and inclination evolution for the binary, Jupiter and Saturn from one random run in each binary model.  We find similar behavior for these larger bodies in all runs of a given model since the mass of the planetesimal disc is not sufficient to significantly affect the binary or giant planet orbits.  The binary orbit remains unchanged but Jupiter and Saturn undergo small oscillations in their eccentricity and inclination.  Immediately we can see that the amplitude of these oscillations increases with binary separation and eccentricity.  The binary perturbations are not enough to destabilize these giant planets, but it does slightly affect the inclination and eccentricity of their orbit.  The ability of giant planets to remain on stable orbits around all the binaries we consider, suggests that circumbinary giant planets are most likely long-lived once formed.

\subsection{Planet formation process}

\begin{figure*}
	\includegraphics[width=1.5\columnwidth]{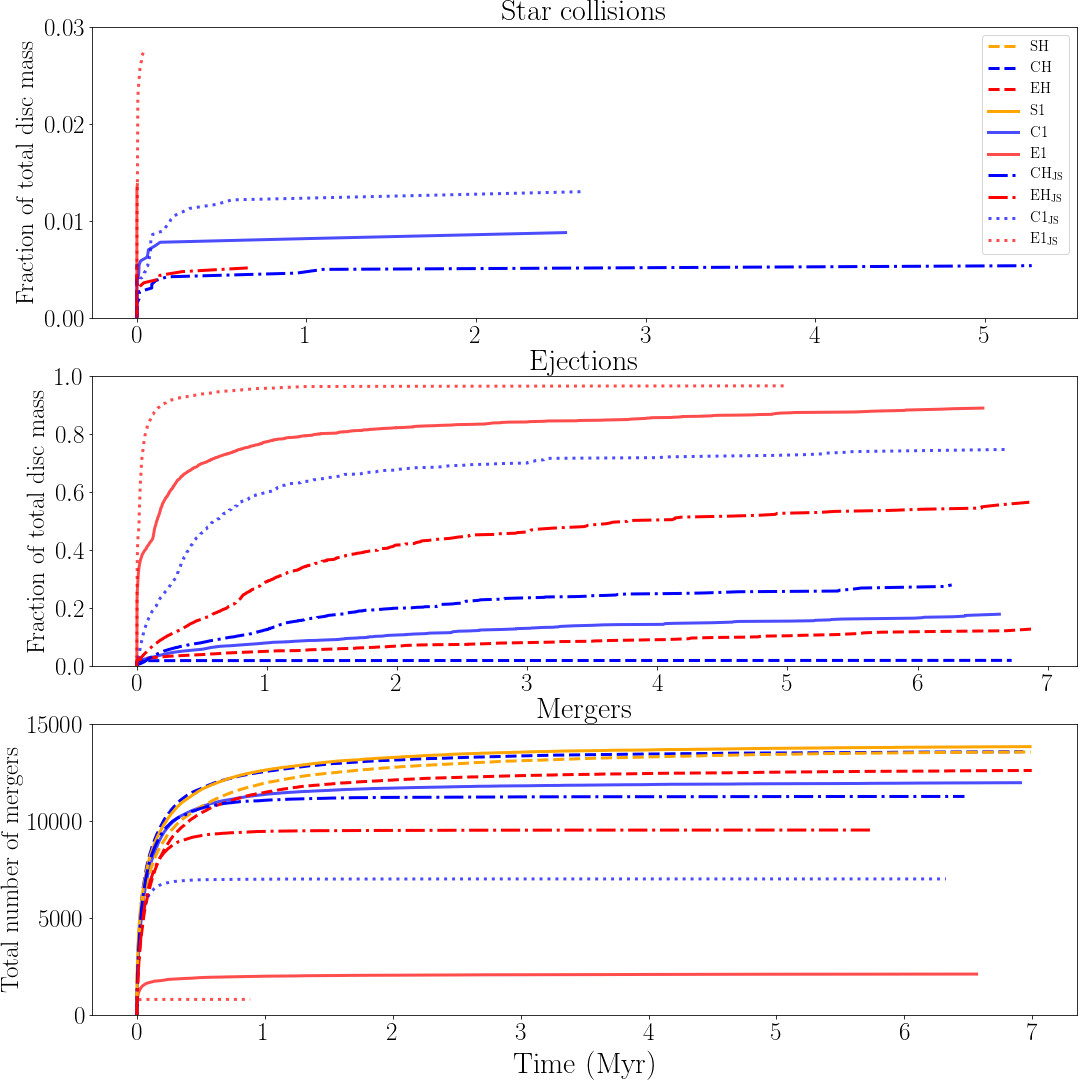}
    \caption{Cumulative number of ejections, star collision and particle mergers for all systems that begin with $4.85 \, M_{\oplus}$ of embryos and planetesimals.  The top panel depicts the cumulative fraction of the total disc mass that collides with one of the stars, the middle panel depicts the cumulative fraction of the total disc mass that is ejected from the system and the bottom panel shows the total number of bodies that merged with a body versus time.  The lines terminate at the time of the last recorded event.}
    \label{fig:ejected_disc_mass}
\end{figure*}

We run our simulations for $7 \,\rm Myr$ which corresponds to a much longer effective timescale for planet formation with the use of an expansion factor of the particle radii as we discussed in Section \ref{nbody}.  As a proxy of how far along the evolution of the system is, Figure \ref{fig:ejected_disc_mass} shows the star collisions, ejections and mergers for all systems as a function of time.  The lines terminate at the time of the last recorded event.  The top panel shows the cumulative fraction of disc mass that has collided with one of the stars.  Star collisions are very infrequent in all systems.  At the onset of the simulations, eccentric binaries experience the highest rate of collisions although these collisions are very short-lived. In the case of the CH model, the systems experience star collisions up to $5 \, \rm Myr$, although at a low rate.  

Disc mass is much more likely to be removed from the system through ejections rather than star collisions.  The middle panel shows the cumulative fraction of the disc mass that has been ejected from the system.  A body is ejected from the system once it exceeds a distance of $100 \, \rm au$.  

The effects of exterior giant planets on terrestrial CBP formation depend on the binary separation and eccentricity.  In general, widely separated binaries have a larger torque than close-in binaries (at a given radius from the centre of mass) and eccentric binaries have a stronger torque than circular binaries.  If the sum of the gravitational perturbations from the binary and giant planets is too large,  the majority of the disc mass is ejected and this  hinders planet formation.  This is the case for binary systems separated by $1 \, \rm au$ that contain Jupiter and Saturn at their current orbits.  The most extreme scenario we consider is the $\rm E1_{\rm JS}$ system which contains a binary with a semi-major axis of $1 \, \rm au$ and $e=0.8$, and also Jupiter and Saturn at their current orbits.  The gravitational perturbations from the binary and giant planets leaves no circumbinary material in the disc to form planets at $7 \, \rm Myr$. However, in systems with giant planets and binaries separated by $0.5 \, \rm au$, ejection rates are moderate. 

We find that binaries separated by $1 \, \rm au$ eject more material than the binaries separated by $0.5 \, \rm au$ because there is a larger fraction of material initially located in $R<a_{\rm c}$ (see Section~\ref{ac}).  Generally, the systems with giant planets eject more material than the systems without giant planets. Systems without giant planets are able to retain more material in their circumbinary discs to grow their planets.

In our single star runs, no mass is ejected from the system and no mass collides with the central star throughout all of the simulations.  All the mass is conserved in these systems as they lack the central perturbations from the binary torque that is expelling mass early on and speeding up planet formation. 

The bottom panel of Figure \ref{fig:ejected_disc_mass} shows the total number of mergers versus time for all systems.  The highest merging rates appear at the beginning of the simulation but some systems are still undergoing steady rates of mergers at $7 \, \rm Myr$.  As expected, we find the total number of mergers and the total number of ejections are inversely related.  The systems with the highest number of ejections ($\rm E1_{\rm JS}$, E1, $\rm EH_{\rm JS}$, $\rm C1_{\rm JS}$), have the lowest number of mergers as there is less material left in the disc to merge.

\subsection{Circumbinary planetary systems}

\begin{table*}
\centering
\caption{Average values and standard deviations for the terrestrial planet multiplicity and planet mass ($M_{\rm p}$), semi-major axis ($a_{\rm p}$), eccentricity ($e$) and inclination ($i$) after $7 \,\rm Myr$ of integration time for all models.  These statistics only consider bodies with a mass larger or equal to $0.1 \, M_{\oplus}$.  We also list the binary separation and eccentricity, and the SPH fit for the initial surface density profiles of each model for reference. Models that include the subscript JS include Jupiter and Saturn, and the model that includes the subscript X begins with a truncated disc that only includes bodies at or exterior to the critical stability limit for an eccentric binary, $a_{\rm c}=3.6 \, a_{\rm b}$.}
\begin{tabular}
{%
  S[table-format=1.0]
  S[table-format=1.2]
  *{2}{ 
    *{2}{S[table-format=1.3(3)]} 
    S[table-format=1.3] 
     S[table-format=1.2]
  }
}
  \toprule
    {Model} & {$a_{\rm b}/ \rm au$} & {$e_{\rm b}$} & {Surface density} &{No. of planets} & {$M_{\rm p}/M_\oplus$} & {$a_{\rm p}/ \rm au$} & {$e$} & {$i^{\circ}$}\\
    \cmidrule(lr){1-9}   
    {SH} &{-}&{-}& {FitCH}&{$5.8 \pm 0.76 $} & {$0.81 \pm 0.60$} & {$2.59 \pm 1.20$} & {$0.06 \pm 0.05$} & {$1.70 \pm 1.53$} \\
    {CH}&{0.5}&{0.0}& {FitCH} & {$5.1 \pm 1.28$} & {$0.89 \pm 0.58$} & {$2.76 \pm 0.88$} & {$0.04 \pm 0.03$} & {$0.96 \pm 0.66$}\\
    {EH}&{0.5}&{0.8}& {FitEH}  & {$3.4 \pm 1.03$} & {$1.22 \pm 0.67$} & {$3.21 \pm 0.79$} & {$0.06 \pm 0.04$} & {$2.63 \pm 2.27$}\\
    {S1}&{-}&{-}& {FitC1}  & {$4.8 \pm 0.75$} & {$0.99 \pm 0.73$} & {$2.98 \pm 0.95$} & {$0.04 \pm 0.04$} & {$1.49 \pm 1.46$}\\  
    {C1}&{1.0}&{0.0}& {FitC1}  & {$2.9 \pm 0.89$} & {$1.35 \pm 0.72$} & {$3.45 \pm 0.68$} & {$0.05 \pm 0.03$} & {$1.35 \pm 0.94$}\\ 
    {E1}&{1.0}&{0.8}& {FitE1}  & {$1.5 \pm 0.57$} & {$0.30 \pm 0.16$} & {$4.10 \pm 0.34$} & {$0.05 \pm 0.03$} & {$1.44 \pm 1.25$}\\   
    \cmidrule(lr){1-9}  
    {$\rm CH_{\rm JS}$}&{0.5}&{0.0}& {FitCH} & {$2.4 \pm 0.91$} & {$1.35 \pm 1.03$} & {$2.26 \pm 0.44$} & {$0.05 \pm 0.04$} & {$1.49 \pm 1.10$}\\
    {$\rm EH_{\rm JS}$}&{0.5}&{0.8}& {FitEH} & {$1.4 \pm 0.69$} & {$1.43 \pm 0.74$} & {$2.62 \pm 0.30$} & {$0.06 \pm 0.05$} & {$3.16 \pm 3.01$}\\
    {$\rm C1_{\rm JS}$}&{1.0}&{0.0}& {FitC1} & {$1.4 \pm 0.61$} & {$0.76 \pm 0.45$} & {$2.85 \pm 0.31$} & {$0.06 \pm 0.02$} & {$1.10 \pm 0.66$}\\ 
    {$\rm E1_{\rm JS}$}&{1.0}&{0.8}& {FitE1}  & {$0.0$} & {$-$} & {$-$} & {$-$} & {$-$}\\ 
    \cmidrule(lr){1-9}   
    {E1$_{\rm X}$} & {1.0} & {0.8}  & {FitE1}  & {$1.4 \pm 0.48$} & {$0.29 \pm 0.13$} & {$4.17 \pm 0.31$} & {$0.05 \pm 0.03$} & {$2.10 \pm 1.42$}\\   
  \bottomrule
\end{tabular}
\label{tab:avg_planets}
\end{table*}

Table \ref{tab:avg_planets} lists the average values of the planet multiplicity, mass, semi-major axis, eccentricity and inclination between all 50 runs for each model after $7 \,\rm Myr$ of integration time.  In the table we only consider bodies with a mass greater than $0.1 \, \rm M_{\oplus}$.  Smaller bodies may still be found in most systems at this time but including these would skew the planet statistics.

\begin{figure*}
	\includegraphics[width=1\columnwidth,height=.75\textheight]{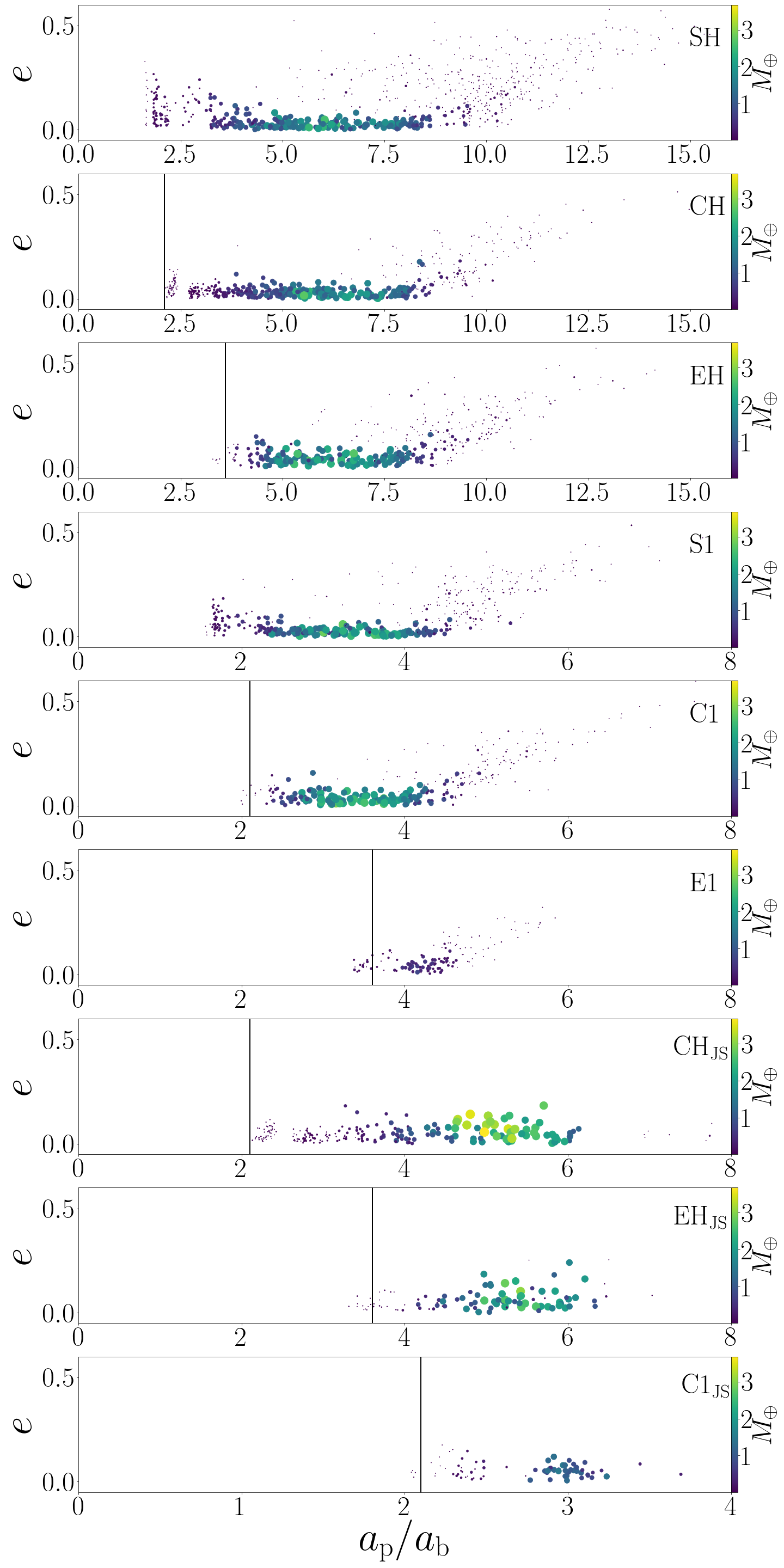}
		\includegraphics[width=1\columnwidth,height=.75\textheight]{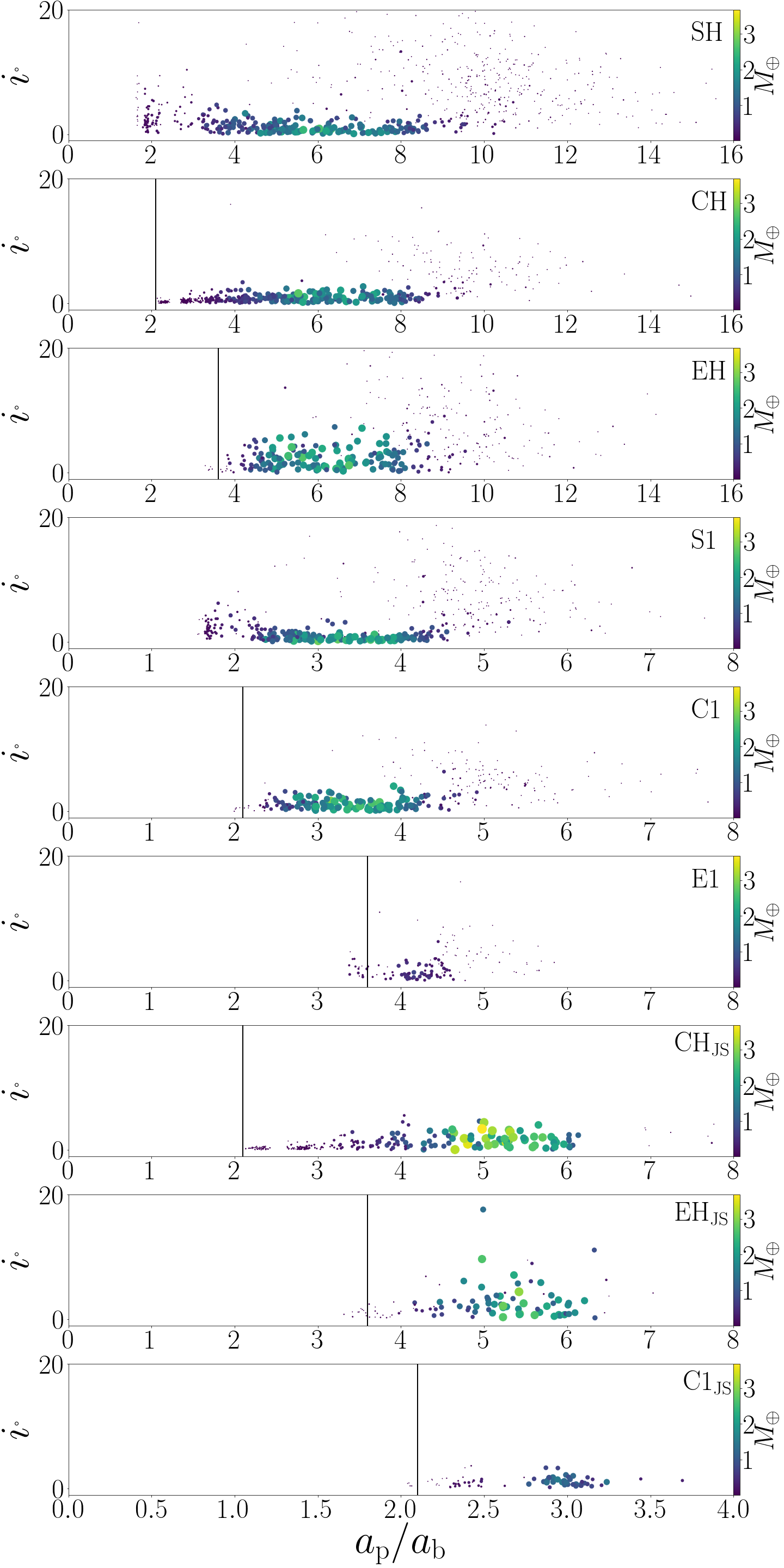}
    \caption{Eccentricity (left panels) and inclination (right panels) vs. the particle semi-major axis, $a_{\rm p}/a_{\rm b}$, for all the bodies that survived $7\, \rm Myr$ of integration time.  The size and colour of the points correspond to the body's mass.} 
    \label{fig:semi_vs_ecc}
\end{figure*}

Figure \ref{fig:semi_vs_ecc} shows the eccentricity (left panels) and inclination (right panels) versus the planet semi-major axis, $a_{\rm p}$, normalised by the binary separation, $a_{\rm b}$, for all the bodies (across all runs) that survived $7\,\rm Myr$ of integration time.
The size and the colour of the particles show the relative masses. We measure the semi-major axis of the bodies in the single star runs in units of their counterpart binary separation for an easier comparison between models.  The black vertical lines mark the critical stability limit, $a_{\rm c}$, for the system.  For all binaries, the planets form exterior to the stability limit although some smaller bodies may be found just interior to the critical radius.  



\subsubsection{Effect of a binary on planet formation}

We first consider the effect of the binary on planet formation. The two single star systems, SH and S1, use the same initial surface density profile as the circular orbit binary simulations CH and C1, respectively. The single star planetary systems and the systems around circular binaries are quite similar but we note a few differences. Larger bodies can be found closer in around single star systems than around binaries.  This is expected as these systems do not contain a central torque. The central torque from a binary speeds up the planet formation process by driving planet-planet interactions, and ejects more material earlier on reducing the reservoir of material available to form terrestrial planets. A population of small mass, high eccentricity and high inclination particles is somewhat depleted by the binary but remain in the single star systems. These effects are comparable to the effects of exterior giant planets on circumstellar systems \citep[see also][]{Childs19}.

The binary star systems form fewer but slightly more massive planets than their single star counterpart.  This indicates that although the planet formation process is happening faster, circular binary systems follow similar planet formation pathways as their circumstellar analogs. The most notable difference in single star systems is that these systems retain more mass.  As a result, the single star systems are able to create higher multiplicity planetary systems than their circumbinary analog.

Widely separated binaries have a larger torque than close-in binaries and so we find that widely separated, circular binaries have fewer but more massive planets than close-in circular binaries.  This is because the larger torque from a widely separated binary speeds up the planet formation process by increasing the rate of mergers. This explains why we see larger planets in the C1 system than the CH system.  However, the stability limit increases with binary separation (see Section~\ref{ac}).  As a result, systems with a larger binary separation are more likely to eject greater amounts of mass.  This is especially true for the discs that hold more mass interior to the critical radii.  A direct consequence of this mass loss are planetary systems with a lower multiplicity.  Our simulation results confirm this as the C1 and E1 runs produce fewer planets than the CH and EH runs.

\subsubsection{Effect of the binary eccentricity}

As the eccentricity of the binary increases, the particle stability limit increases (see Section~\ref{ac}). Thus, the range of semi-major axes with stable orbits decreases. This increase in unstable disc regions greatly decreases the efficiency of planet formation by ejecting the majority of the disc mass in the inner system.  Consequently, we find more planets around circular binaries than around eccentric binaries and they are able to form closer in.  Comparing CH to EH, we see that planets that form around eccentric binaries may form with higher mass, eccentricity and inclination.



 The widely-separated eccentric binaries found in our E1 runs have the largest torque of all the systems we consider and produce the fewest and smallest planets.  However, compared to their circular counterpart, the perturbations from a close-in eccentric binary seems to be a \textit{sweet spot} for planet formation as these systems produce more massive planets on average with a relatively low mass ejection rate.

\subsubsection{Effect of giant planets}

Comparing the $\rm CH_{\rm JS}$ and $\rm EH_{\rm JS}$ systems to CH and EH, respectively, we see that the systems with giant planets produce fewer but more massive planets. Because the central torque for the binaries separated by $0.5 \, \rm au$ is relatively small, the additional exterior perturbations from the giant planets aids planet formation.

The large regions of unstable space in systems with giant planets are evident in Figure \ref{fig:semi_vs_ecc}.  We see that giant planets efficiently truncate the outer edge of the planetesimal and embryo disc around $3-3.5 \, \rm au$.  This truncation only permits planets to form between the stability limit $a_{\rm c}$ and about $3.5 \, \rm au$. This means that in the wider orbit binaries we consider, planet formation is largely inhibited. There are no planets in E1$_{\rm JS}$ and only a few small planets around  C1$_{\rm JS}$.  It should also be noted that the large secular resonances from Jupiter and Saturn occur interior to this outer stability limit such as the $\nu_6$ resonance that is around $2\,\rm au$ in the solar system \citep[e.g.][]{Froeschle1986,Morbidelli1991}. Giant planets are efficient at removing the large population of low mass, high eccentricity and high inclination bodies seen to be most populous in the single star systems but present in all the simulations without giant planets.

Because no planets formed in the $\rm E1_{\rm JS}$ runs, we suggest that terrestrial planets are very unlikely around widely separated ($a_{\rm b} \geq 1 \, \rm au$), highly eccentric, coplanar binaries that harbor giant planets. The addition of giant planets into the already inefficient systems around widely-separated, eccentric binaries removes almost all likelihood of planet formation.

\subsection{Expansion factor convergence tests}
\label{ef}

\begin{figure*}
	\includegraphics[width=1.6\columnwidth]{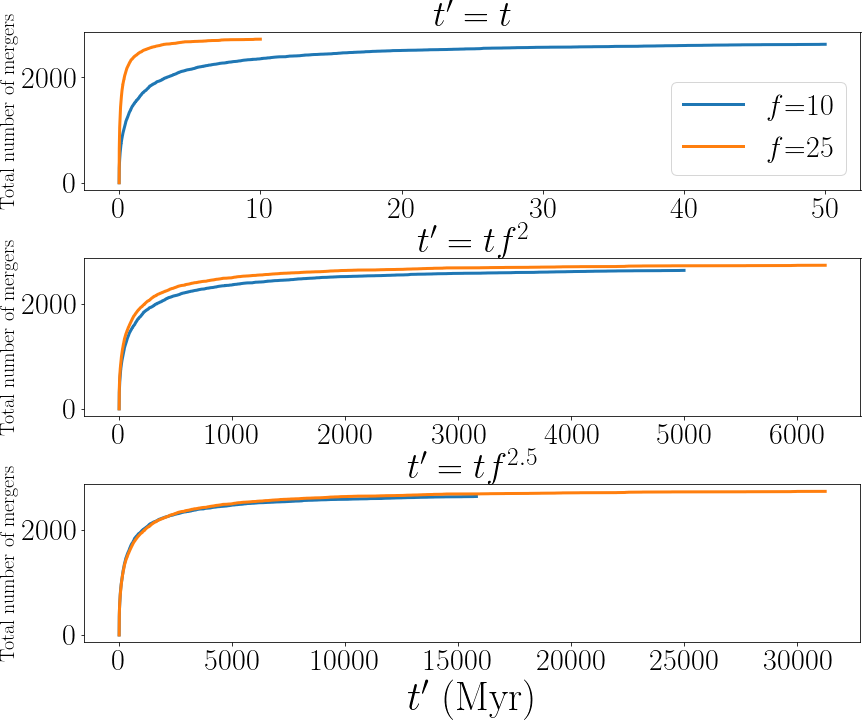}
    \caption{Total number of mergers versus the effective time, $t'$.  The top panel shows the total number of mergers versus simulation time $t$ with no scaling.  The middle and lower panels show this same data but multiply the simulation time $t$ by $f^2$ and $f^{2.5}$. We find $f^{2.5}$ is a more accurate scaling for predicting the \textit{effective} time of planet formation than $f^{2}$ when perfect merging is assumed.}
    \label{fig:ef_tprime}
\end{figure*}

\begin{table*}
\centering
\caption{We consider the SH and CH models with different numerical integrators and expansion factors $f$.  We list the model, integrator, expansion factor, time and the resulting terrestrial planet multiplicity, planet mass ($M_{\rm p}$), semi-major axis ($a_{\rm p}$), eccentricity ($e$) and inclination ($i$).  In the CH systems, we evaluate all runs at $7 \rm Myr$.  In the SH models, we evaluate the systems with $f=25$ at $1 \, \rm {Myr}$ of simulation time, and the runs with $f=10$ at $10 \, \rm Myr$ since these are similar \textit{effective} times. 
These statistics only consider bodies with a mass $\geq 0.1 \, M_{\oplus}$ and the data from 10 runs for each setup.}
\begin{tabular}
{%
  S[table-format=1.0]
  S[table-format=1.2]
  *{2}{ 
    *{2}{S[table-format=1.3(3)]} 
    S[table-format=1.3] 
     S[table-format=1.2]
  }
}
  \toprule
{Model}&{Integrator}&{$f$} & {Time/Myr} &{No. of planets} & {$M_{\rm p}/M_\oplus$} & {$a_{\rm p}/ \rm au$} & {$e$} & {$i^{\circ}$}\\
\cmidrule(lr){1-9}   
{CH}&{\texttt{IAS15}}&{50}&{7}& {$6.0 \pm 0.67$} & {$0.92 \pm 0.56$} & {$2.75 \pm 0.80$} & {$0.03 \pm 0.02$} & {$0.82 \pm 0.47$}\\   
 {CH}&{\texttt{IAS15}}&{25}&{7}& {$5.1 \pm 1.28$} & {$0.89 \pm 0.58$} & {$2.76 \pm 0.88$} & {$0.04 \pm 0.03$} & {$0.96 \pm 0.66$}\\ 

      \cmidrule(lr){1-9} 
      
{SH}&{\texttt{IAS15}}&{25}&{$1$}&{$7.1 \pm 0.70 $} & {$0.59 \pm 0.45$} & {$2.78 \pm 1.09$} & {$0.06 \pm 0.05$} & {$1.38 \pm 1.70$}\\ 

{SH}&{\texttt{Mercurius}}&{25}&{$1$}& {$7.7 \pm 1.49$} & {$0.56 \pm 0.46$} & {$2.82 \pm 1.08$} & {$0.05 \pm 0.03$} & {$1.52 \pm 1.49$}\\  
      
{SH}&{\texttt{Mercurius}}&{10}&{$10$}& {$6.2 \pm 1.25$} & {$0.68 \pm 0.54$} & {$2.87 \pm 1.35$} & {$0.09 \pm 0.06$} & {$3.23 \pm 2.87$}\\

  \bottomrule
\end{tabular}
\label{tab:expansion_factors}
\end{table*}

To check how the expansion factor affects our simulations, we first compare the binary simulation CH with a higher expansion factor. Then we consider the single star model SH with a lower expansion factor since modeling only a single star allows us to use a faster numerical integrator.

\subsubsection{Higher expansion factor}

First we experiment with a larger expansion factor to see if the results converge.  We run ten CH models with $f=50$ and compare the results to ten CH runs with $f=25$.  We compare the systems at $7 \, \rm Myr$ which corresponds to evolution timescales much greater than what is needed for the systems to fully evolve.  We list the resulting planetary systems in Table \ref{tab:expansion_factors} and find that both expansion factors produce similar systems suggesting that expansion factors larger than 25 may be suitable for similar $n$-body studies.

We emphasize that the focus of this study is not to accurately predict final planet properties but to identify differences in terrestrial circumbinary planet formation trends as a function of binary separation and eccentricity.

\subsubsection{Single star model with lower expansion factor}

We now perform convergence tests with a lower expansion factor using the SH model.  Because the SH model uses only one star, we are able to use a faster hybrid integrator which is not well suited for binary studies.  \texttt{Mercurius} is a hybrid of the high accuracy non-symplectic integrator \texttt{IAS15} and the symplectic integrator \texttt{WHFAST} \citep{Rein2019}. To study how the integrator affects the simulation results we first include 10 runs of the SH systems with $f=25$ and the \texttt{IAS15} integrator and compare to 10 runs with the \texttt{Mercurius} integrator.  In Table \ref{tab:expansion_factors} we see that the two integrators for the SH model with $f=25$ produce similar systems however the \texttt{IAS15} integrator returns slightly fewer but higher mass planets on more eccentric and inclined orbits suggesting that \texttt{IAS15} more accurately captures the effects of gravitational scattering than \texttt{Mercurius}.

Using the \texttt{Mercurius} integrator and the SH model we compare ten runs with an expansion factor of $f=10$ to ten runs with $f=25$.  Because a smaller expansion factor corresponds to a longer effective timescale we integrate the systems with $f=10$ for $50 \rm \, Myr$. 
In order to compare the SH systems with $f=10$ to the SH systems with $f=25$ at the same \textit{effective} time, we first need to determine how $f$ reduces the evolution timescale.  To do this, we use the total number of mergers as a proxy of the effective time, which we denote by $t'$.  Figure \ref{fig:ef_tprime} shows the total number of mergers across all ten runs with $f=10$ and $f=25$ versus the simulation time $t$ without any scaling, and also with $t$ scaled by $f^{2}$ and $f^{2.5}$.  Although \citet{Kokubo_2002} suggest that the evolution timescale is reduced by $f^{2}$, we find that the evolution timescale is more accurately reduced by $f^{2.5}$ when perfect merging is assumed.

To check if the $f=10$ systems converge to the $f=25$ runs, we evaluate the planet properties at the same effective time $t'=tf^{2.5}$.  We evaluate the $f=25$ runs at $1 \, \rm {Myr}$ and the $f=10$ runs at $10 \, \rm {Myr}$. 
Table \ref{tab:expansion_factors} lists the integrator and expansion factor used, the integration time the system is evaluated at and the average values and standard deviations of the resulting planetary system multiplicities, planet mass, semi-major axis, eccentricity and inclination for the bodies with a mass $\geq \, 0.1 \, \rm M_{\oplus}$.  

Both systems produce planets with similar semi-major axes, but the $f=10$ systems produce slightly fewer and more massive planets that are on more eccentric and inclined orbits than the $f=25$ systems. 
Because $f=25$ systems merge bodies together more quickly than the $f=10$ systems, the bodies do not have enough time for the orbits to grow to excited states via planet-planet scattering which lowers the probability of collisions.  This explains why the $f=25$ systems slightly underestimate the planet eccentricity, inclination and mass.

\subsection{Effect of particles that begin in unstable regions}
\label{unstable}

We first consider directly the effect of the initially unstable particles that may form at orbital radii $R<a_{\rm c}$ in the gas disc since it is uncertain whether planetesimals can form so close to the binary. We choose the simulation with the most mass initially interior to $a_{\rm c}$, model E1, and we run the same simulation but remove the particles that are initially in $R<a_{\rm c}$.
We perform 50 runs for the E1$_{\rm X}$ model.  The X in the subscript of a model name means the inner radius of the disc has been truncated at the critical particle stability limit.  We use the same disc setup around the eccentric binaries as described in Section \ref{nbody} however, we only include bodies with an initial semi-major axis greater than or equal to the critical stability limit for a binary with $e=0.8$, that is $a_{\rm c}=3.6 \, a_{\rm b}$. Thus, the disc in E1$_{\rm X}$ does not begin with the same amount of material as E$1$.

The bottom row in Table \ref{tab:avg_planets} shows the average values and standard deviations for the terrestrial planet multiplicity, planet mass ($M_{\rm p}$), semi-major axis ($a_{\rm p}$), eccentricity ($e$) and inclination ($i$), after $7 \,\rm Myr$ of integration time.  These statistics only consider bodies with a mass larger or equal to $0.1 \, M_{\oplus}$.
Comparing the results from model with a truncated discs to the model without a truncated disc we find that the results are very similar.  The E1$_{\rm X}$ system initially contains 145 bodies, including 12 embryos, yielding a total planetesimal and embryo mass of $2.35 \, M_{\oplus}$ which is $\sim 49 \%$ of the  solid disc mass in model E$1$.  Even with only less than half of the solid disc mass available, the E1$_{\rm X}$ simulation results in almost identical systems as the E1 system. This suggests that the early outward scattering of the unstable bodies interior to $a_{\rm c}$ does not significantly alter the evolution of the planetary system. 

There is a  second initially unstable region at the outer edge of our particle disc in the simulations for which we include the giant planets.  We expect that particles that begin in this region similarly have little effect on the formation of the terrestrial planets. The particles are rapidly ejected at the start of the simulation. As we discussed in Section~\ref{hydro}, we did not use an initial surface density profile for the particles motivated by the shape of a gap in the gas disc carved by Jupiter. The gap size is estimated to be down to about $3.65\,\rm au$ while our simulations extend to $4\,\rm au$.  In all of the simulations that include the giant planets, with the exception of one C1$_{\rm JS}$ run, there are very few planetesimals ($m=0.0093\,  \rm M_{\oplus}$) and only one embryo ($m=0.093\,  \rm M_{\oplus}$) found exterior to $3.65\,\rm au$ after $7 \, \rm Myr$. Across all of the C1$_{\rm JS}$ simulations, one planetesimal is found exterior to the gap edge and also a planet with $0.32 \,  \rm M_{\oplus}$ at $3.69 \, \rm au$.  The embryo for this planet began just interior to the gap edge at $3.57 \, \rm au$ and migrated outwards slightly in time. Excepting this particular, there are no planets with a mass greater than $0.1\,\rm M_\oplus$ in $R>3.44\,\rm au$. Thus, the sharp truncation of the initial particle disc at the outer edge does not affect the outcomes of the simulations.

\section{Conclusions}
\label{concs}

Using $n$-body simulations, we have modeled the late stages of CBP formation around various binary systems.  We used the results of hydrodynamic gas disc simulations to determine the initial distribution of Moon and Mars-sized bodies for our $n$-body simulations.  We considered both eccentric ($e$=0.8) and circular ($e$=0) binary orbits with a circumbinary disc of planetesimals and embryos.  Some of our runs included Saturn and Jupiter at their current orbit.  We also simulated a subset of runs using only a single star to disentangle the effects of the binary and a  run that begins only with bodies at or exterior to the critical particle stability limit to explore the effects of initially unstable particles that may form on stable orbits in the gas disc.

To conclude, we list our main findings here:
\begin{itemize}

    \item A central binary strongly affects the initial distribution of particles available for terrestrial planet formation. The gas disc extends closer to the binary than the critical particle stability limit. Solid bodies form on orbits that are unstable once the gas disc has dissipated and are quickly ejected.  The outward scattering of the bodies does not significantly alter the evolution of the bodies found on stable orbits.
    
    \item 
    The CBP formation process around close circular binaries  ($a_{\rm b}\lesssim 0.5\,\rm au$) is very similar to the circumstellar planet formation process. However, the torque from the binary speeds up the planet formation process by promoting body-body interactions and driving the ejection of planet building material. This leads to slightly fewer but more massive planets around a close binary.
    A sufficiently wide binary provides a large central torque which can prevent terrestrial planet formation.
    
     \item Eccentric binaries can eject large amounts of disc material 
     and form fewer terrestrial planets than circular binaries. The wider and more eccentric the binary, the more mass that is ejected from the terrestrial planet forming region. However, around a close eccentric binary, these planets are more massive, more eccentric and more highly inclined than around a circular orbit binary.

   \item Giant planets reduce the range of stable orbits for planets to form and systems with giant planets form fewer terrestrial planets. The combined perturbations from giant planets and the binary torque can destroy planet formation completely for a wide and/or eccentric binary. However, the few planets formed around close binaries with giant planets have larger mass, larger eccentricity and higher inclination than the planets in systems without giant planets.

    \item The giant planets remain on stable orbits in all of our simulations suggesting that circumbinary giant planetary systems can be long-lived once formed.
\end{itemize}

\section*{Data availability statement}
The SPH simulations results in this paper can be reproduced using the {\sc phantom} code (Astrophysics Source Code Library identifier {\tt ascl.net/1709.002}). The $n$-body simulation results can be reproduced with the {\sc rebound} code (Astrophysics Source Code Library identifier {\tt ascl.net/1110.016}).  The data underlying this article will be shared on reasonable request to the corresponding author.   

\section*{Acknowledgements}

We thank an anonymous referee for useful comments that have improved the manuscript.  We thank Daniel Tamayo and Hanno Rein for helpful discussions. Computer support was provided by UNLV’s National Supercomputing Center. AC acknowledges support from a UNLV graduate assistantship and from the NSF through grant AST-1910955.  Simulations in this paper made use of the REBOUND code which can be downloaded freely at \url{http://github.com/hannorein/rebound}. RGM acknowledges support from NASA through grant 80NSSC21K0395.



\bibliographystyle{mnras}
\bibliography{ref}









\bsp	
\label{lastpage}
\end{document}